\documentclass[useAMS,usenatbib]{mn2e}
\usepackage[pdftex]{graphicx}
\usepackage{times}
\usepackage{gensymb}
\usepackage[authoryear]{natbib}

\title[Orbital instability of close-in exomoons in non-coplanar systems]{Orbital instability of close-in exomoons in non-coplanar systems}
\author[ Y.-C.~Hong, M.~S.~Tiscareno, P.~D.~Nicholson, and J.~I.~Lunine]{Yu-Cian Hong$^{1}$\thanks{E-mail:
ycsylva@gmail.com (YCH)} , Matthew S. Tiscareno$^{2}$, Philip~D.~Nicholson$^1$, and Jonathan I. Lunine$^{1,2}$\\
$^{1}$Astronomy Department, Space Sciences Building, Cornell University, Ithaca, NY 14853, USA\\
$^{2}$Center for Radiophysics and Space Research, Space Sciences Building, Cornell University, Ithaca, NY 14853, USA}

\begin{document}

\date{Accepted 2015 Feb 11}

\pagerange{\pageref{firstpage}--\pageref{lastpage}} \pubyear{2015}

\maketitle

\label{firstpage}

\begin{abstract}
This work shows the dynamical instability that
 can happen to close-in satellites when planet oblateness is not accounted for in non-coplanar multiplanet systems. 
Simulations include two secularly interacting Jupiter-mass planets 
mutually inclined by 10$\degree$, with the host planet either oblate or spherical.  With a spherical host planet, moons within a critical planetocentric distance experience high inclinations and in some cases high eccentricities, while more 
distant moons orbit stably with low inclinations and eccentricities, as expected.  These counter-intuitive dynamical phenomena 
disappear with an oblate host planet, in which case the moons' Laplace plane transitions from the host planet's equatorial plane
 to the host planet's precessing orbital plane as their semi-major axes increase, and all moons are dynamically
 stable with very mild changes in orbits.  Direct perturbation from the perturbing planet has been investigated and ruled out as an explanation for the 
 behavior of the innermost satellites, therefore leaving the central star's perturbation as the cause.
 Instability occurs while the nodal precession of the satellite and the central star (as seen from the host planet's frame) approaches the 1:1 secular resonance.  In non-coplanar systems,
around a non-oblate planet, the nodal precession of the moon becomes slow and comparable to that of the planet, giving rise to resonant configurations.  The above effect needs to be taken into account
in setting up numerical simulations.

\end{abstract}


\section{Introduction}

  Giant planets in our Solar System all have extensive natural satellite systems, which has led to extensive studies on whether exoplanets likely
 also harbor satellites.  Exomoons could be interestingly diverse, habitable, or provide constraints on planet formation and evolution theory.  Since the discovery of 
exoplanets, the habitability of their moons has been studied considering different aspects such as
 stellar illumination, stellar irradiation, satellite atmosphere, planet magnetic field, etc. \citep{heller12,heller13b,heller13a,kaltenegger,williams97}. 
Exomoons have some advantages over exoplanets on the extent of habitable orbital configurations.
 The possibilities of them 1) orbiting around the confirmed giant planets in the classical liquid water habitable zone
determined by stellar illumination, planetary greenhouse effect and others \citep{kasting}, and 2) retaining
 the right temperature for stable liquid water outside the classical habitable zone by tidal heating \citep{heller13b,reynolds,scharf}
could make them desirable targets for Earth-like habitability studies.

Various techniques have been studied for detection of exomoons.  1) transit timing variation of the moon-hosting close-in planets
 \citep{agol,holman,kipping09a,kipping12,sartoretti,simon} can detect moons down to 0.2 Earth size with the Kepler mission \citep{kipping09c}.  2) Microlensing 
detection of exomoons can reach down to 0.01 Earth masses \citep{bennett96,bennett02,han02,han08}.  It is better suited to detect moons that orbit planets
 distant from the parent star.   A sub-Earth mass exomoon candidate has been found orbiting a free-floating giant planet via this technique \citep{bennett14}. 
3)  Direct imaging can detect bright, tidally heated exomoons down to 1 Earth size with temperatures above 300K and 600K with JWST and Warm Spitzer \citep{peters}.

However, questions arise on whether exomoons exist because we understand very little about their formation and evolution processes.
  Sources of perturbation to their orbital evolution are very different from within the Solar System. Unlike the rather circular,
 co-planar, and well-separated orbits of Solar System planets, the orbits of many exoplanets are eccentric and they may also be mutually
 inclined as generally predicted by the planet-planet scattering model \citep{chatterjee,juric,marzari,raymond}. Some 
multiplanet systems are found to be relatively compact, and many exoplanets are on extremely close orbits around the parent star. Competition 
between perturbations to their orbits and the host planet's gravitational attraction determine whether moons can remain on stable orbits.

Previous works have explored satellite orbital stability in exoplanet systems in different dynamical settings.  Some have focused on
 systematical numerical studies of single planet systems \citep{barnes,domingos,donnison,holman99}, in which the satellite removal 
processes could be stellar tidal stripping, violation of the Hill stability criterion, etc.  Recent studies proceeded
 to multiple planet systems, where planetary perturbation on the moons becomes very important because orbital spacings between planets are compact 
or they experience planetary close encounters \citep{frouard,gong,hong12,nesvorny,payne}.  Planet oblateness was considered
 as negligible in the scope of the above mentioned works.

This work will show that, in multiplanet systems where the orbits of planets are mutually inclined, close-in satellites
situated within a critial planetocentric distance where perturbation to the orbit is usually dominated by planet oblateness \citep{nicholson} can become dynamically unstable when planet oblateness is neglected.
Non-coplanarity provides a path for the occurence of the nodal precession of the planets and satellites, and as will be shown in section 4, the approach of secular resonances of the 
nodal precession rates of the innermost satellites with the host planet ( or the central star in the host planet's frame) may be the cause for instability.
Such instability needs to be considered in setting the satellite-hosting planet's oblateness in order to obtain reasonable numerical results.

We simulate systems with two planets mutually inclined by 10$\degree$.  The host planet is treated as spherical in simulation set 1
 and as oblate in simulation set 2 for comparison.  As will be shown in section 3, in simulation set 1, satellites
 with small planetocentric distances gain high inclination or leave planet-bound orbits, while distant satellites stay
 on low inclinations and relatively unperturbed orbits.  By contrast, such unusual dynamical phenomena for the close-in satellites disappear in simulations with an oblate host planet, now that the gravitational potential
in the region close to the planet becomes dominated by the $J_2$ moment term.

\section{SIMULATION SETTING}
This work uses the N-body symplectic integrator Mercury \citep{chambers}.  All simulations use the Bulirsch-Stoer algorithm, the most accurate
 for simulating bodies that perturb each other closely, though the slowest in the package, in order to exclude integration error as a cause for the 
problematic orbits of satellites in this work.
Unless otherwise specified, the simulations are configured as follows : 1) the integration error limit in orbital energy and angular momentum is $ 10^{-12}$, 
2) the integration time-step is 0.1 days, in order to accurately integrate satellites with orbital periods as short as a few days, and 3) the
simulation duration is 1 million years.

Simulations test the orbital stability of primordial satellites in non-coplanar two-planet systems.  All simulations consist of a central star, two giant 
planets and their satellites.  The scene is set after the final stage of satellite formation and after the circumplanetary disk has dissipated. 
 The mass and radius of the central star are about that of the Sun $\mbox{--}$ 1 Solar mass and 0.0046AU.  The masses and densities of the giant planets
 resemble those of Jupiter $\mbox{--}$ $9.548 \times 10^{-4}$ Solar mass and 1.3 $g/cm^3$.  The inner planet is 
always at 5 AU from the star, and the orbits of both planets are nearly circular and mutually inclined by 10$\degree$.  The satellites are massless
 test particles.  They orbit around the inner planet (hereafter the host planet) at a range of planetocentric semi-major axes $a_s$ = 0.008 
$\mbox{--}$ 0.25 Hill radii (hereafter $R_H$)\footnote{In the circular restriced 3-body problem, the Hill radius of a planet marks the boundary 
within which the gravitational attraction of the planet dominates the star's tidal field. $R_{Hill} = a_p \left(\frac{m_p}{m_*}\right)^{\frac{1}{3}}$, where
 $m_p$ is planet mass and $m_*$ the central star's mass.}, or $a_s$ = 0.002731 $\mbox{--}$ 0.08534 AU.  
The innermost satellite approximates Io's position around Jupiter and lies beyond 3 times the Roche limit for 
satellites with a density higher than 0.5 $g/cm^3$, so that it's safe to neglect tidal mass loss or tidal disintegration \citep{guillochon}. The outermost satellite is well 
within the stability limit \footnote{This is the stability limit for prograde satellites in single planet systems with the planet and satellites on nearly circular orbits.} 
 of $\sim$ 0.5 $R_H$ \citep{domingos}, as we have no interest in satellites that become unstable due to violation of the Hill stability criterion.  Most of the simulations have 3 satellites
 at $a_s$ = 0.008, 0.06, and 0.25 $R_H$, unless specified.  The discussion section gives some special focus on the satellite at $a_s = 0.008\,R_H$, hereafter called the innermost satellite.  
The satellites are on initially circular ($e < 0.0001$), prograde orbits, and are co-planar with the host planet's initial orbital
 and equatorial plane.  The gravitational interactions between satellites are not included in any simulations.  All bodies are non-spinning. 

In total there are 4 sets of simulations.  

Simulation set 1 has 25 simulations and the only variable between simulations is the semi-major axis of the exterior planet 
(hereafter the perturber), which spans the range $a_p$ = 6.602 $\mbox{--}$ 12.612 AU, corresponding to planet-planet orbital period ratios 
$\frac{P_2}{P_1}$ = 1.517 $\mbox{--}$ 4.01, 
and mutual Hill radii of 3.21 $\mbox{--}$ 4.01.  Due to our interest in stable planetary systems, perturbers in all simulations initially lie outside the boundary of 
global chaos. \citep{veras}
 If the planet separations were inside the global chaos limit, they would be bound to undergo gravitational scattering. The dynamical instability criterion associated with
 non-coplanar systems will be further discussed in section 3.    

Unlike what is assumed in simulation set 1, the majority of real giant planets may be oblate.  Giant planets are non-rigid bodies that contain spin angular momentum inherited from the
 orbital angular momentum of the smaller bodies in the circumstellar disk that form them, so they experience rotational flattening.  A planet 
distorted in shape modifies the gravitational potential around itself\footnote{Assuming the planet 
has uniform density and is an ellipse with axial 
symmetry, the modification can be approximated by a term with the second order Legendre polynomial with its factor the $J_2$ moment, 
$V (r,\theta) = -\frac{Gm}{r} [ 1 - J_2 \cdot (\frac{R}{r})^2 \cdot P_2(cos\theta) ]$ \citep{murray}. G is the gravitational constant, m the planet mass,
 R the planet's equatorial radius, r the planetocentric distance, $\theta$ the angle from the principle axis,
 and $J_2$ a dimensionless constant.}.

Simulation set 2 is a duplicate of simulation set 1, except that the host planet is oblate, with a
 $J_2$ moment of 0.0147, equal to that of Jupiter, with its equatorial bulge fixed in the initial reference plane.  The difference between simulation sets 1 and 2 leads 
to results that address the main theme of this work: the host planet's $J_2$ moment plays a crucial role in the stability of close-in satellites $\mbox{--}$ that the absence of it can lead to
instability or even loss of satellites.

Simulation set 3 is a duplicate of simulation set 1, except that perturber-moon interaction is turned off.  This set tests whether the perturber directly causes 
the counter-intuitive behavior of the innermost satellites.

Simulation set 4 takes the simulation with $\frac{P_2}{P_1}$ = 4.01 from 
simulation set 1, and changes the integration error limit and time-step to $10^{-13}$ and 0.001 days.  Simulation set 4 double checks whether the  problematic 
orbits of satellites in simulation set 1 could be
due to the lack of integration accuracy.

\section{SIMULATION RESULTS}

 All output orbital elements of planets in this section are calculated in the star-centered frame, and those of the satellites in the host-planet-centered
 frame.  Different planes of reference are used where appropriate, and they are 1) the initial orbital plane of the host planet and satellites, which
 is identical to the host planet's equatorial plane, 2) the host planet's precessing orbital plane, and 3) the
system's invariable plane determined by the total angular momentum.

\subsection{Planet Orbits}

Starting with a 10$\degree$ mutual inclination and nearly circular orbits, 
the planets in simulation set 1 and 2 are initially well-separated enough so their mutual perturbations cause little change in semi-major axes and eccentricities
over 1 million years, and their mutual inclination remains close to constant at all times.

\begin{figure}
  \centering
    \centering 
   \includegraphics[width=0.99\columnwidth]{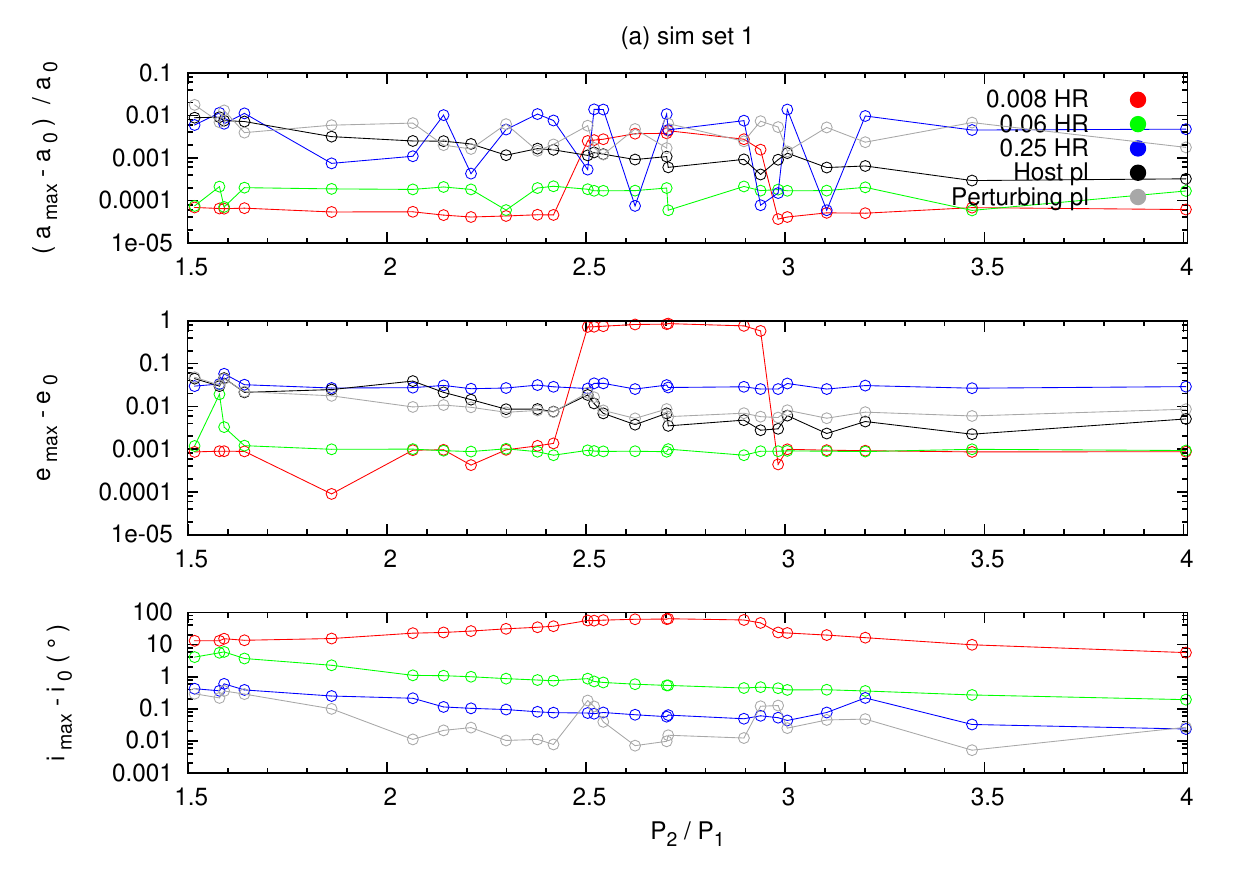}
   \centering
   \includegraphics[width=0.99\columnwidth]{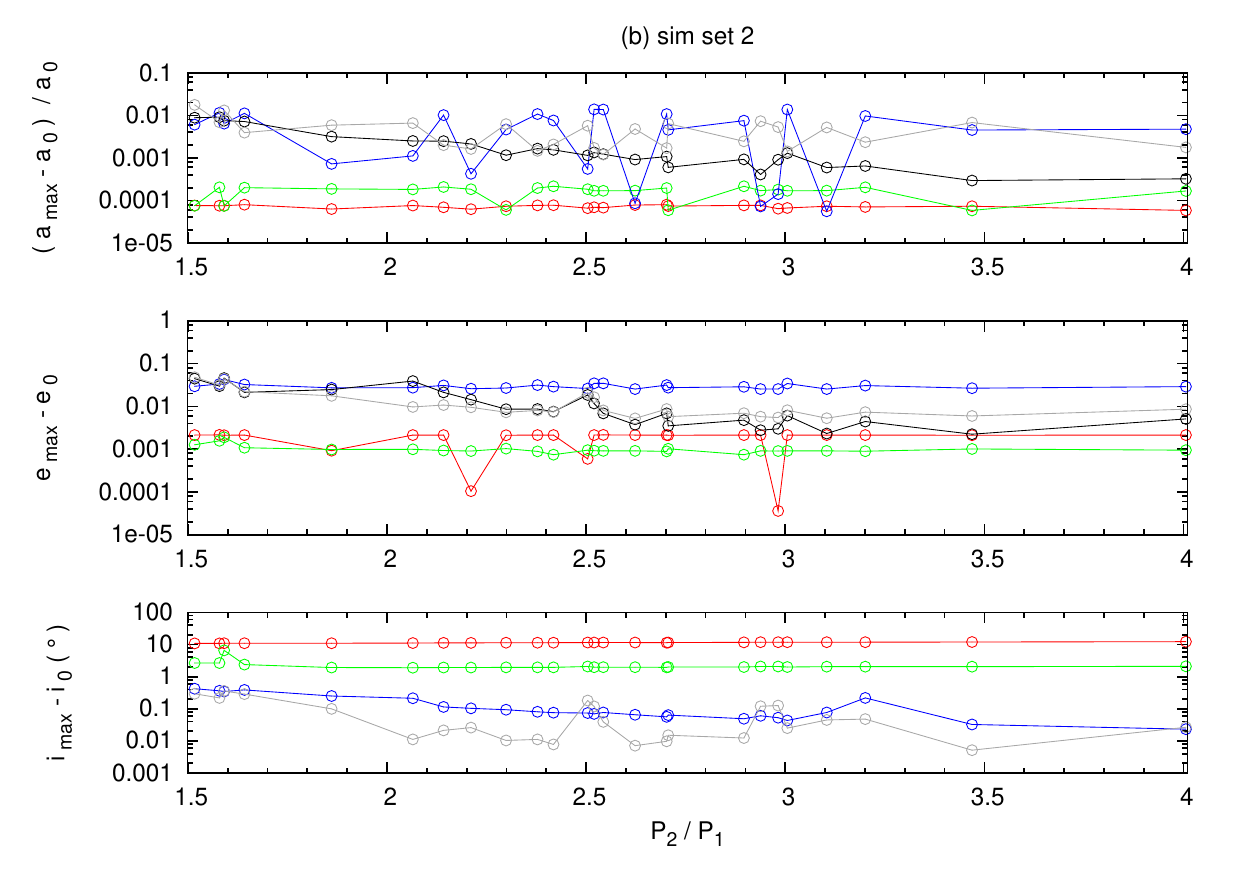} 
  \caption{%
  The maximum change in the orbital elements (semi-major axis $a$, eccentricity $e$, and inclination $i$) of planets and satellites throughout each simulation in set 1 and 2.  
The reference plane for both simulation sets is the host planet's precessing orbital plane.
The legend box shows the color representation of planets, and satellites by their semi-major axes.
Planets in both simulation sets experience very mild changes in orbits, and the maximum deviation in both simulation sets
 are somewhat similar.  Like the planets, change in semi-major axes of the satellites are mild.  However, in simulation set 1, there is a trend for the innermost satellites to experience higher inclinations 
than the outer ones.  The innermost satellites in 
the range $\frac{P_2}{P_1}$ = 2.50 $\mbox{--}$ 2.94 even reached inclinations beyond 40 $\degree$ and eccentricities above 0.5.  On the other hand, the more distant satellites
at $a_s$ = 0.06 and 0.25 $R_H$ experience much milder changes.  In simulation set 2, the trend for distant satellites to stay close to the host planet's orbital plane is 
the same, but the misbehaved innermost satellites now find a home in the host planet's equatorial plane; the innermost and outer satellites' eccentricities
 now both remain small, and all satellites are stable.}\label{fig1}
\end{figure}

However, as seen from the initial reference plane, the inclinations of the two planets
 oscillate sinusoidally with a constant period but different amplitudes
\footnote
{This is as predicted by the Laplace-Lagrange theory of secular perturbation.  The range of their inclination can be approximated by:
$0\leq I_{1}(t) \leq \frac{2I_0}{1+ \sqrt{\frac{a_1}{a_2}}}$, and $ I_0 \times \frac{1-\sqrt{\frac{a_1}{a_2}}}{1 + \sqrt{\frac{a_1}{a_2}}} \leq I_2(t) \leq I_0$ \citep{veras}. 
The suffix 0 represents initial values at t = 0, suffix 1 represents the inner planet (host planet), and suffix 2 represents the outer planet (perturber).}.  The direction of the 
host planet's equatorial bulge, which is fixed on the initial reference plane, in turn oscillates sinusoidally with the same period and angle as the host planet's inclination 
oscillation when seen from the host planet's orbital plane.  This setting could give rise to unphysical motion for close-in satellites since in reality the planet's obliquity
should evolve as its orbital inclination changes; however, it does not affect the simulation results within this work since the innermost satellites stay on the planet's equatorial
plane, as they should be around planets with low obliquity. \citep{goldreich}
The maximum changes in semi-major axes of the planets 
in all simulations of set 1 and 2 are under 2$\%$, in eccentricities under 0.05, and in inclinations under 0.4$\degree$, using the host planet's
 orbital plane as reference.  Dynamical instability does not occur in these selected simulations, and their orbits deviate from each other little enough for the purpose of comparing satellite orbits between the 
2 simulation sets. (figure \ref{fig1}).

As a side note, in a few simulations, planet orbits evolve chaotically.  Non-coplanarity allows the planets with separations larger than the global chaos
 limit to undergo gravitational scattering (orbit crossing) or have their orbital elements evolve significantly.  These simulations have their initial 
planet separations close to the boundary of global chaos or to first- or second-order resonances \citep{veras}.
We do not include those simulations in the discussion but only focus on the stable ones, so that the effect of planet oblateness plays the major factor in 2 comparable sets
of simulations.

\subsection{Satellite Orbits}

In simulation set 1, the absence of the host planet's oblateness perturbation causes the innermost satellites to undergo unusual orbits, as will be illustrated
in the sub-sections below.  However, by treating the host planet as oblate in simulation set 2, moon orbits become tame and stable, and their motion is predicted to a good 
approximation by the Laplace theory.

\begin{figure*}
  \centering
   \includegraphics{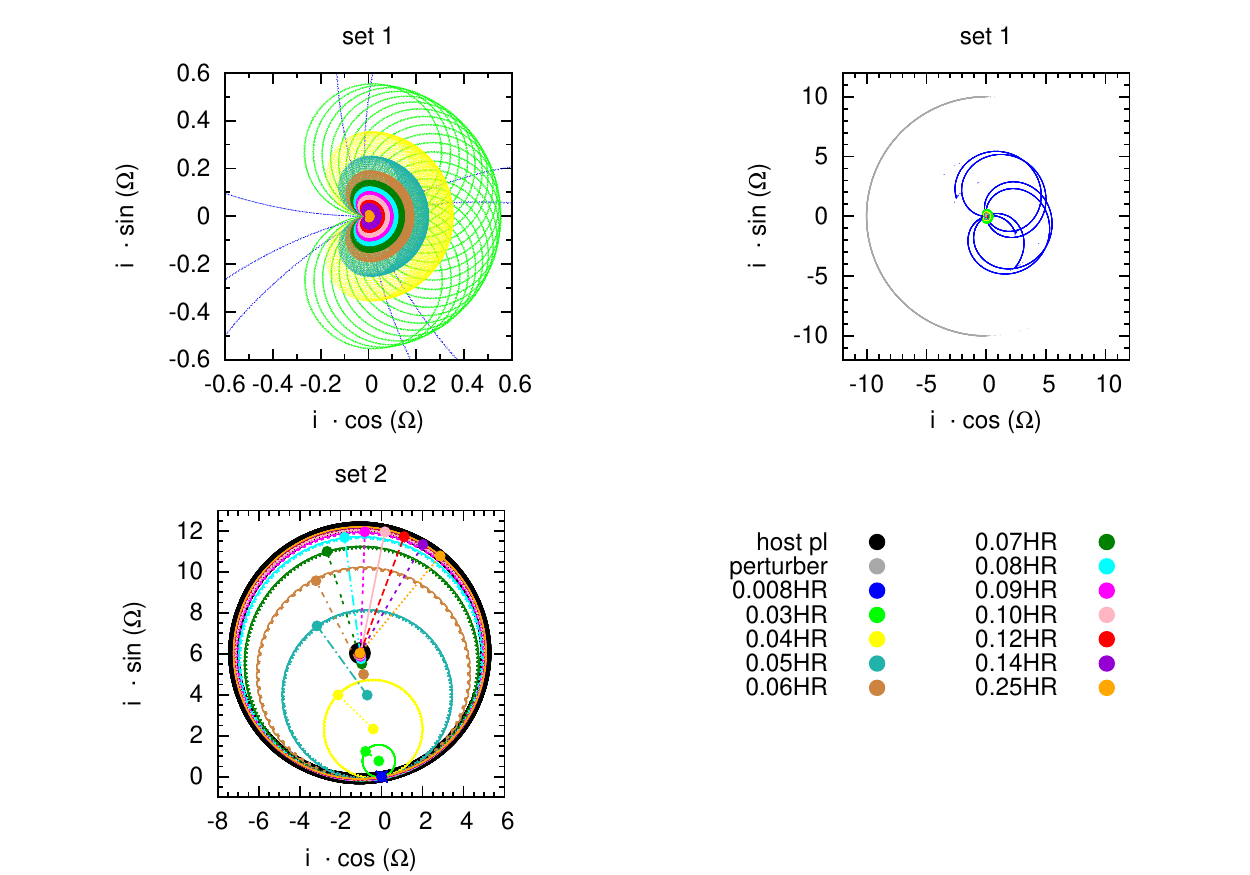}
  \caption{ The evolution of the angular momentum axes of planets and 12 satellites in the simulation with $\frac{P_2}{P_1}$ = 4.01 in both 
simulation set 1 and 2, in the first 229,090 years. The legend box shows color representation of planets, and satellites by their semi-major axes.  The host planet's plot
is exaggerated in thickness for clarity.
The reference plane for simulation set 1 is the host planet's precessing orbital plane, and for simulation set 2 the host planet's equatorial plane.
Simulation set 1 is plotted twice with different scales.  Distance from the origin represents inclination, which increases with decreasing
satellite semi-major axes.  Each satellite appears to precess around some plane that itself is precessing around the host planet.
On the other hand, satellites in simulation set 2 all precess around an average plane of their own ( the center of each circle), which moves from the
host planet's equatorial plane to its orbital plane as the satellites' semi-major axes increase.
The radius of the circle is their inclination relative to the average plane.  The innermost satellite only stays on the host planet's
equatorial plane, while the outer most ones stay close to the host planet's orbital plane.}\label{fig2}
\end{figure*}

\subsubsection{Satellite orbital plane}

In the upper panel of figure \ref{fig2} is a selected simulation from set 1 with $\frac{P_2}{P_1}$ = 4.01, where satellites with smaller semi-major axes are more inclined 
than those with larger ones, and their orbital normal, the angular momentum axis determined by inclinations and nodes, precesses around a plane that itself is precessing 
around the host planet's orbital normal, which sits at the origin of the plot.  On the other hand, satellites with larger semi-major axes
tend to stay close to the host planet's orbital plane, or in other words, the time evolution of their angular momentum axes follow that of the host planet closely.
  
When this simulation is rerun in set 2 with an oblate host planet,
 the satellites become well-behaved.  As shown in the lower panel of figure \ref{fig2}, the innermost satellites stay on the host planet's
equatorial plane at all times, and satellites at different semi-major axes orbit around their individual Laplace 
planes, which transition from the host planet's equatorial plane to the host planet's orbital plane as $a_s$ increases.  No satellites have gone onto a plane outside this 
region, as do the innermost satellites in simulation set 1.  Because the perturber's  gravitational influence is negligible compared with that of the star, as is shown in figure \ref{fig2}
, the satellites' motion appears to conform with the 200-year-old theory of Laplace surface \citep{tremaine09}.

\subsubsection{High inclination and instability}

\begin{figure}
  \centering
    \centering 
   \includegraphics[width=0.99\columnwidth]{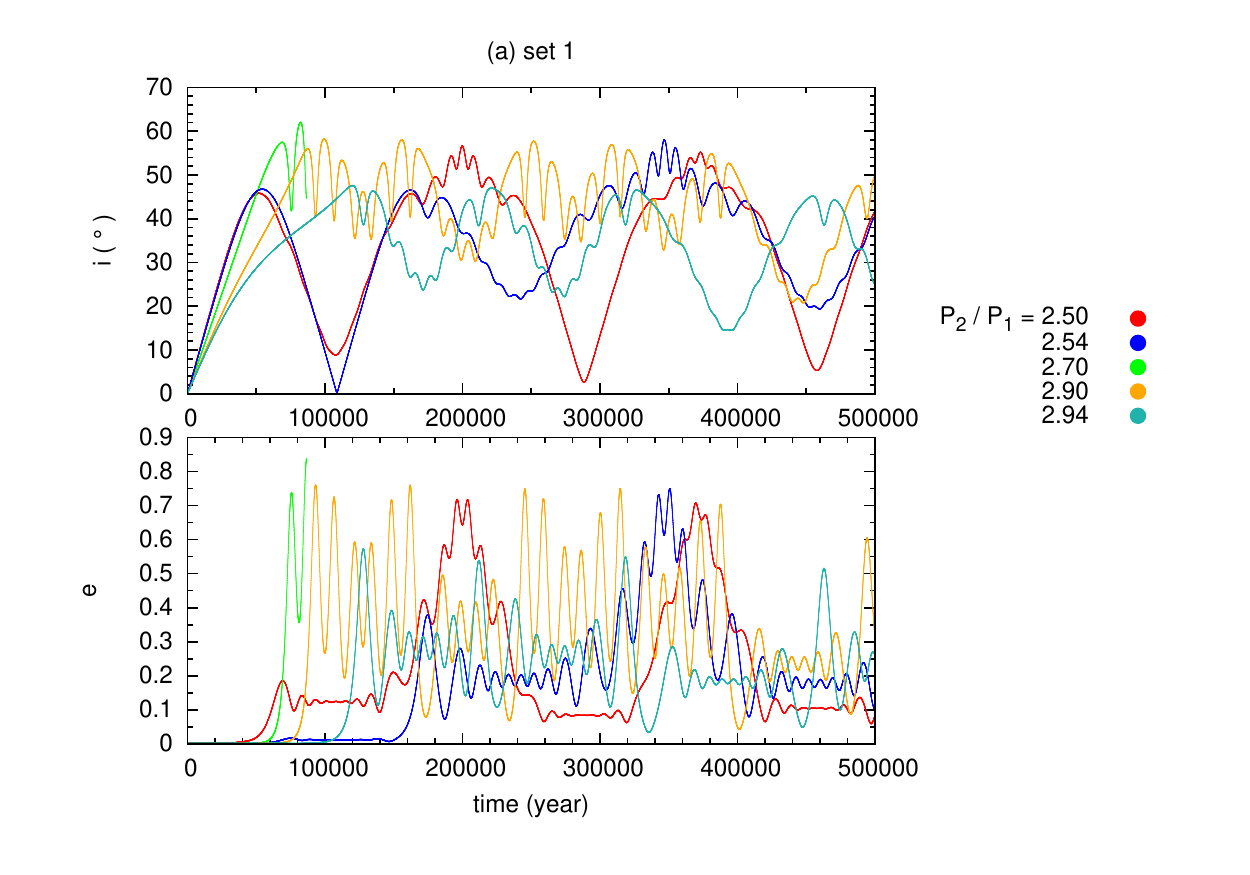}
   \centering
   \includegraphics[width=0.99\columnwidth]{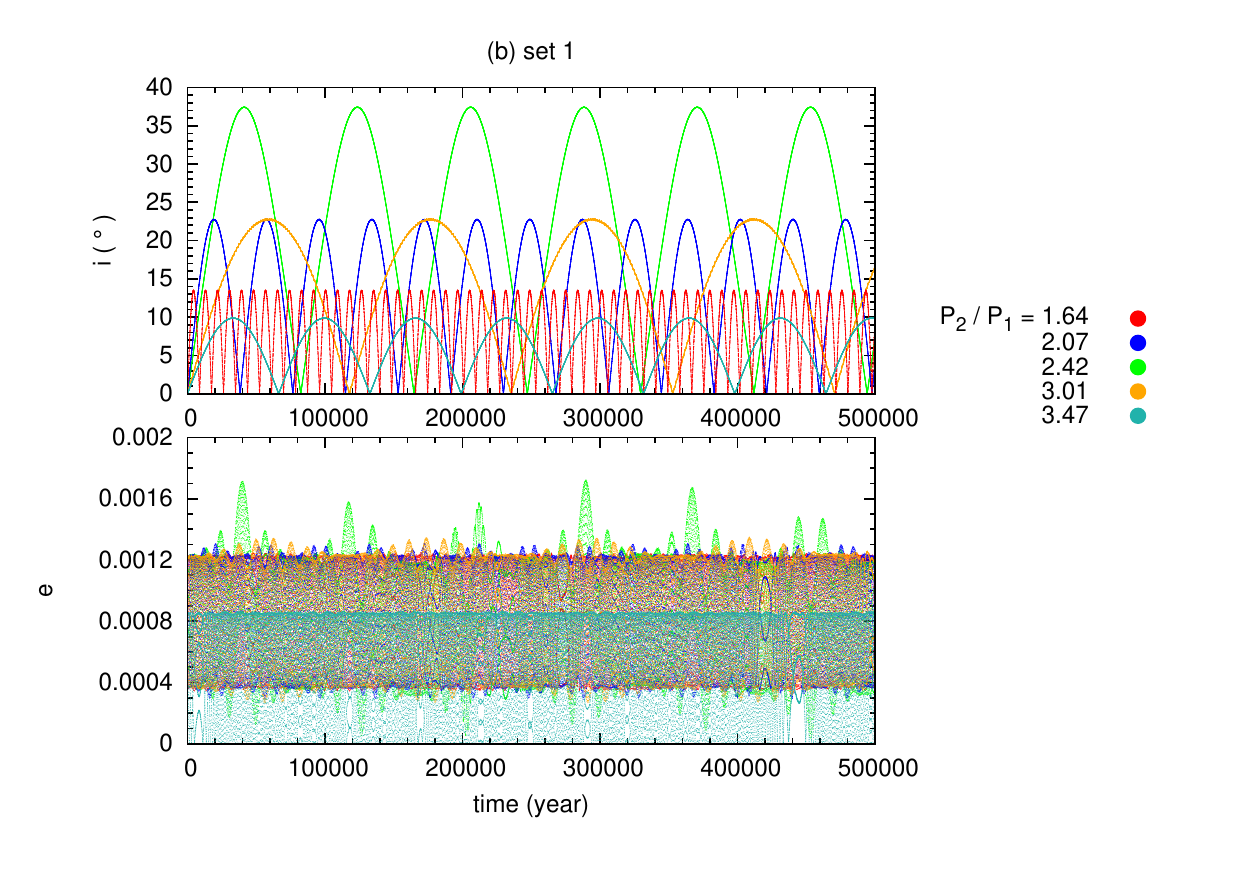}
  \caption{Time evolution of the innermost satellites' ( $a_s$ = 0.008 $R_H$) orbital elements in several simulations within simulation set 1.  The reference plane is the host planet's precessing orbital plane.  
The legend box shows color representation of satellites by the simulation they belong to.  The evolution of semi-major axis is not shown because the change is very mild ( $<$ 1$\%$).  
In (a) the innermost satellites in $\frac{P_2}{P_1}$ = 2.50 $\mbox{--}$ 2.94 undergo rather disturbed orbits in inclinations and eccentricities.
  All of them have reached i $>$ 40 $\degree$ and most of them have e $>$ 0.7.  The innermost satellite in $\frac{P_2}{P_1}$ = 2.70 has the highest e ( $>$ 0.8) and
 i ( $>$ 60 $\degree$), and is lost to collision to the host planet due to its high e.
(b) shows the orbits of the innermost satellites in $\frac{P_2}{P_1}$ outside the range 2.50 $\mbox{--}$ 2.94.  Their inclinations are milder ( $<$ 40 $\degree$),
although still higher than the outer moons,
 and much less disturbed.  Their eccentricities are also very small ( $<$ 0.002).  
The peak inclination drops as $\frac{P_2}{P_1}$ moves away from the region 2.50 $\mbox{--}$  2.94.
}\label{fig3}
\end{figure}

As shown in figure \ref{fig1}, simulation set 1 exhibits an overall trend for close-in satellites to stay farther away from the host planet's orbital plane than more distant ones in both stable and unstable cases.
The innermost satellites' maximum inclination varies from 5.6$\degree$ to 64.1$\degree$.  In each simulation with $\frac{P_2}{P_1}$ = 2.50 - 2.94, the innermost satellite's
 maximum inclinations are higher than 47$\degree$, and eccentricities above 0.58, whereas in other simulations of set 1 inclinations are under 40$\degree$ and eccentricities
are negligibly small.  On the other hand, more distant satellites expriences much milder changes in orbits.  Figure \ref{fig3} illustrates how the innermost satellites' inclinations and eccentricities evolve through time.
The innermost satellite in $\frac{P_2}{P_1}$ = 2.70 is lost to collision with the host planet due to its high eccentricity ( $>$ 0.8 ).  The reason for the satellites' eccentricity to rise might be that the increased 
inclinations cause some perturbative terms to become large. 
In the lower panel, for simulations outside the region $\frac{P_2}{P_1}$ = 2.50 - 2.94, satellite orbits are much less disturbed and instability doesn't occur, 
although most innermost satellites are still more inclined than the outer ones;
 also, the peak inclinations of moons drop as $\frac{P_2}{P_1}$ moves away from the unstable region, and all moons have eccentricities under 0.002.
As a side note, though instability occurs in simulation set 1, all satellites have fractional changes in semi-major axis under the order of 0.01.  This points to secular effects
as the culprit, as they are well known to alter eccentricities and inclinations while doing no work on the semi-major axes.

By contrast, in simulation set 2, as shown in figure \ref{fig1} (b), the innermost 
satellites' maximum changes in eccentricity are all under 0.005, and most of them have zero inclination relative to the the host planet's equatorial plane at all times, with only 
two reaching $i \sim$ 0.02$\degree$.  All satellites, including outer ones, stay on stable orbits around their host planet, and the change in orbital
 elements of all satellites in the simulation set remains reasonably mild throughout.

The motion of the satellites in simulation set 2 are more physical and stable compared with simulation set 1.  It will lead to significant differences
in important metrics such as exomoon survival rate.  As the setting in simulation
 set 2 is also more realistic, we conclude that adding the host planet's equatorial perturbation is a necessary measure for simulating exomoons in systems 
with mutually inclined planets.

\section{Potential causes for instability}

\subsection{Direct perturbation by perturber}

Simulation set 3 tests the hypothesis that the inner moons' behavior in set 1 is caused by the direct gravitational effects of the perturbing planet, which
 may compete with the direct effect of the star (which is also inclined, due to the perturbing planet's effect on the host planet) such that the star's 
effects dominate for most satellites but fall off with decreasing satellite semi-major axis more steeply than do the direct effects of the perturbing planet, thus 
leading to abnormal behavior by the innermost satellites.

\begin{figure}
  \centering
   \includegraphics[width=0.99\columnwidth]{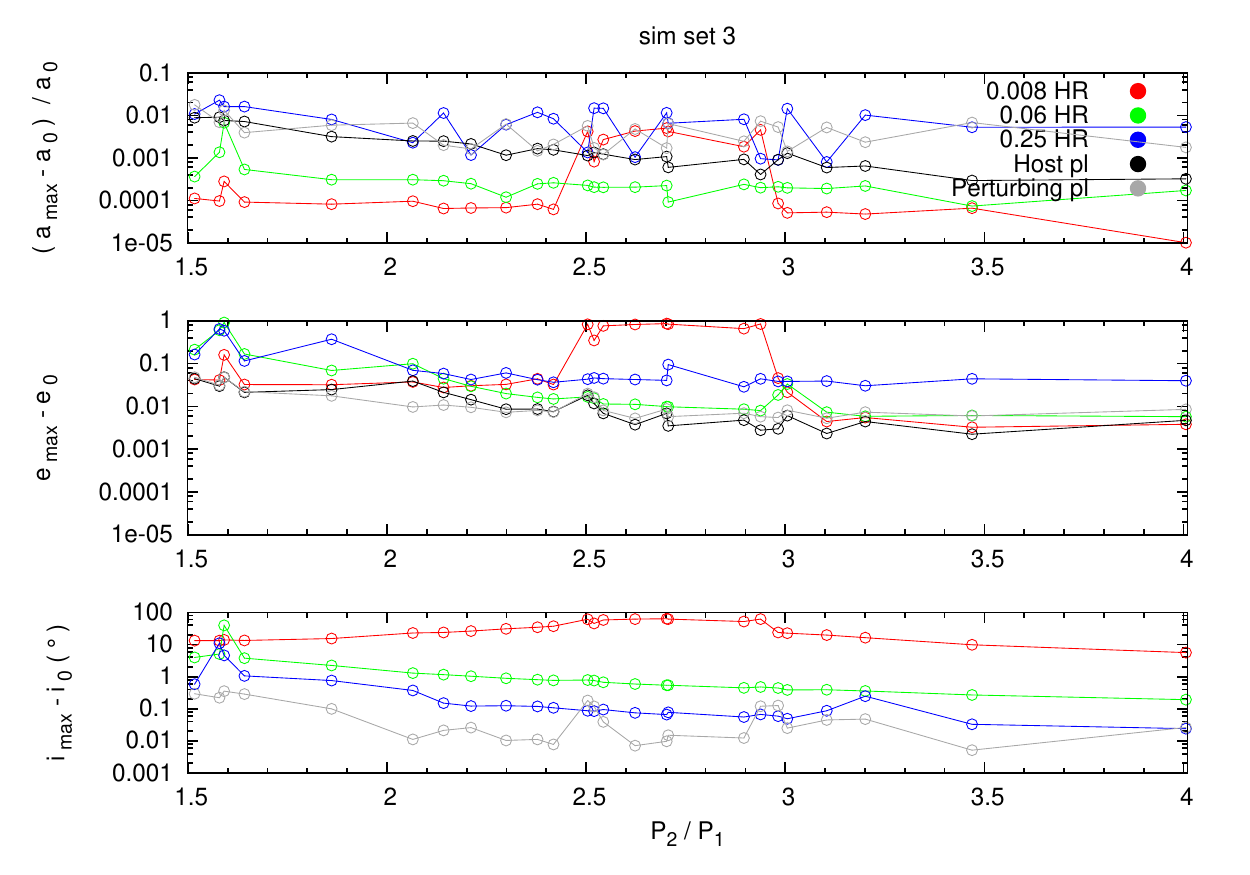}
 
  \caption{%
 The maximum change in the orbital elements of planets and satellites throughout each simulation in simulation set 3.  
The reference plane is the host planet's precessing orbital plane.  Planet orbits resemble that of set 1, and satellites demonstrate similar behavior as
 in figure \ref{fig1} (a), although many have eccentricities $\sim$ 1 order of magnitude higher.}\label{fig4}
\end{figure}

Simulation set 3 is a duplicate of simulation set 1, except that perturber-moon interaction is turned off.  The unusual behavior of the satellites will
disappear if the above hypothesis is right.  The planets' orbits do not differ from set 1 and 2,
 and the moons' behaviors are qualitatively the same as in set 1. The evolution of the angular momentum axes of moons in $\frac{P_2}{P_1}$ = 4.01 
in set 3 is almost the same as in set 1 (upper panels of figure \ref{fig2}).
Also, as shown in figure \ref{fig4}, inner moons have higher inclinations than outer moons.  The innermost moons in $\frac{P_2}{P_1}$ = 2.50 - 2.94 
 have inclinations above 45.8$\degree$ and eccentricities mostly above 0.65, whereas in the rest of the simulations their inclinations are below 40$\degree$ and
 eccentricities mostly below 0.04 but one at $e$ = 0.11.  Many of the moons' maximum changes in eccentricities are at least an order of magnitude higher than in set 1.  The highereccentricities in set 3 appear to be a result of forced perturbation by the eccentric star.  By comparison, in set 1, the direct influence of the perturbing planet could give the satellites enough free eccentricity to avoid such forced perturbation.  Despite this difference, simulation set 3 demonstrates that direct perturbation from the perturber is not the major cause for the satellites' behavior in set 1.  Instead, the culprit may involve the precession of the host planet's orbital plane (or, from the host planet's
point of view, the star's orbital plane) indirectly caused by the perturber.

\subsection{Secular resonance}

Around an oblate planet, the critical distance separates the region where planet oblateness and stellar perturbation dominates.  It is located where the two perturbations are equal.
\begin{equation}
	a_{crit} = \left(2\,J_2\,R_p^2\,a_p^3\,\frac{m_p}{m_*}\right) ^{\frac{1}{5}},
\end{equation}
where $J_2$ is the quadruple moment of the planet, $R_p$ the planet's radius, $a_p$ the planet's distance from the star, $m_p$ the planet's mass, and $m_*$ the star's mass. \citep{kinoshita}
The host planet's $a_{crit}$ in set 2 is 0.015 AU $\left(\,0.044\,R_H\,\right)$.  The fact that the innermost satellite in set 1 lies within this distance but with the planet oblateness missing could potentially give rise to instability associated with
secular resonances.

When a satellite's nodal precessional motion resonates with one of the system's eigenfrequencies, it is said to be in secular resonance.  Secular resonance of nodes can increase the inclination of a body.  Examples in the
Solar System are asteroids and trojans that resonate with $\nu_{16}$, which roughly equals Jupiter's nodal precession rate.  

\begin{figure}
  \centering
   \includegraphics[width=0.6\columnwidth, angle = 270]{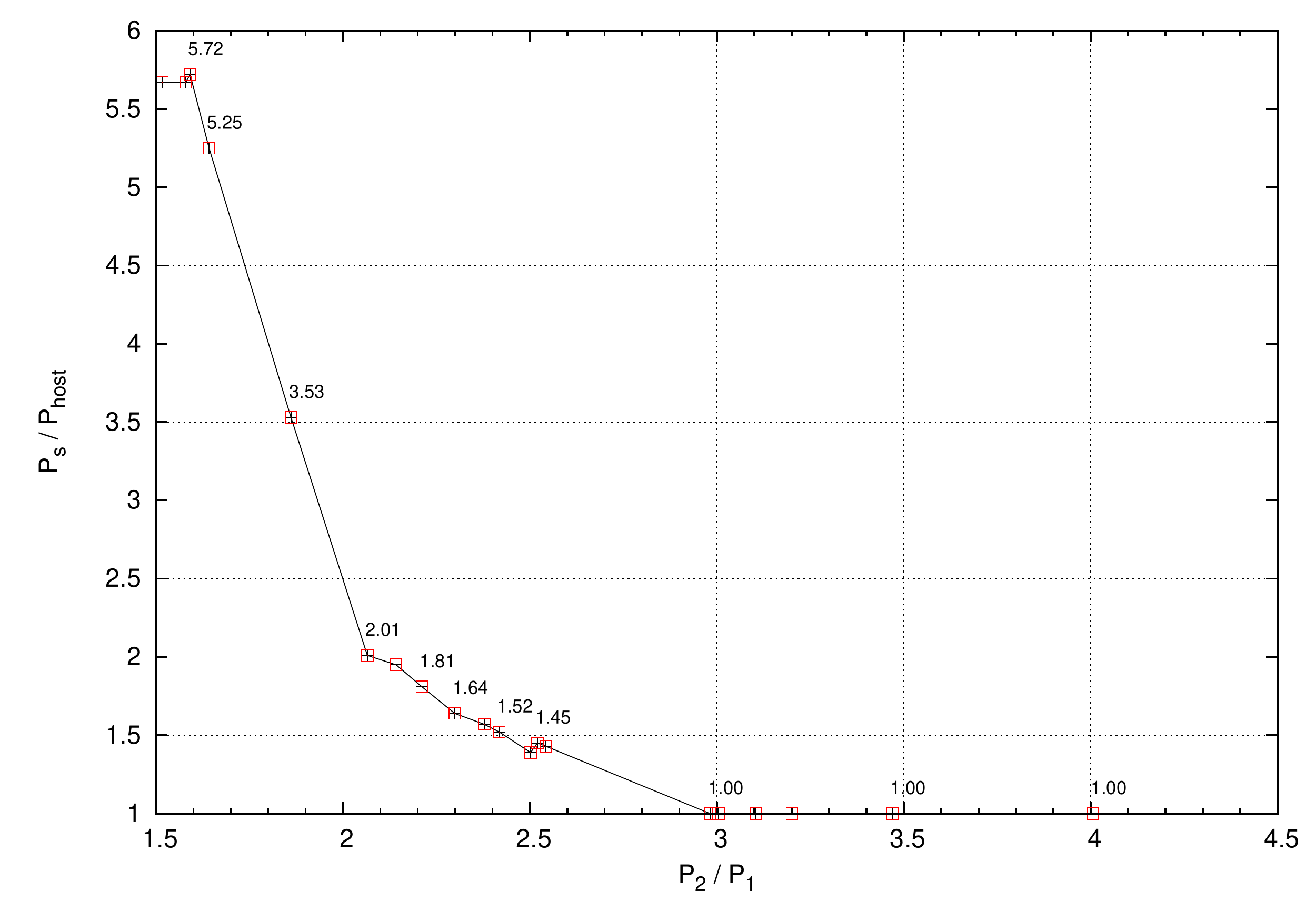}
  \caption{%
The vertical axis represents the period ratio of nodal precession of the innermost satellite versus that of the host planet ( or the star in the host planet's frame), and on the horizontal axis is the planet's orbital period ratio.  The reference plane is the system's invariable plane.  At small $\frac{P_2}{P_1}$, the host planet precesses much faster than the satellite, but its precession rate decreases much faster than the satellite with increasing $\frac{P_2}{P_1}$, therefore the former eventually hit the 2:1, 3:2, and then enter the 1:1 resonance with the latter.  As they enter the 3:2 resonance, the effect of the approach of the 1:1 resonance starts to destabilize the satellites.  The instability could be associated with the chaotic zone
in the vicinity of the 1:1 secular resonance or in the overlap region of the 3:2 and 1:1 resonances.
}\label{fig5}
\end{figure}

The precessional motion of the satellites in this work is made possible by the mutual inclination of the planets, which left the initial orbital plane after time t = 0.  Due to the lack of planetary bulge, the satellites' nodal precession is dictated by stellar perturbation.  The nodal precession rate is given by
\begin{equation}
        -\,\frac{3}{4}\,\frac{n_*^2}{n_s}\,cos\,i\,,
\end{equation}
where $n_*$ is the mean motion of the star and $n_s$ that of the satellite, and $i$ is relative to the host planet's equatorial plane.
Satellites with smaller orbital distances precess slower than distant ones, and the calculated period of one precession cycle is $\sim 3.6 \times 10^{4}$ years for the innermost satellite.  In the simulations, the innermost satellite's period of nodal precession is $3.7 \times 10^{4}$ years at $\frac{P_2}{P_1}$ = 1.52, and it slowly increases as $\frac{P_2}{P_1}$ increase, probably due to an increase in $i$.  On the other hand, the nodal precession period of the host planet is $6.4 \times 10^3$ years at $\frac{P_2}{P_1}$ = 1.52, much shorter and increases much faster than the innermost satellite as $\frac{P_2}{P_1}$ increases (fig. \ref{fig5}), since the major source of perturbation to the satellite is the central star and to the host planet is the perturbing planet.  This allows the host planet and the satellite to hit several resonances at different $\frac{P_2}{P_1}$.  As shown in figure \ref{fig5}, instability happens as they hit the 3:2 and upon entering the 1:1 secular resonance ($\frac{P_2}{P_1}$ = 2.50 - 2.94), and the 1:1 resonance as seen from the invariable plane ($\frac{P_2}{P_1} >$ 2.94) appears to have a stabilizing effect.  The fact that unstable cases all lies in between the 3:2 and 1:1 secular resonances could point to resonance overlap as the cause for instability.  \footnote{In perturbed Hamiltonian systems, the separatrices on the phase plane are not perfect thin lines but disarrayed layers.
As the perturbative term is being cranked up, the upper and lower separatrices of two largest resonances spread toward each other, and as the critical perturbation is reached, the separatrices touch, and then systems
originally locked in a isolated resonance would transition through the chaotic zone weaved by the layered separatrices and experience chaos.  This is called Chirikov's resonance overlap criterion. \citep{chirikov}}

\begin{figure}
  \centering
    \centering 
   \includegraphics[width=0.99\columnwidth]{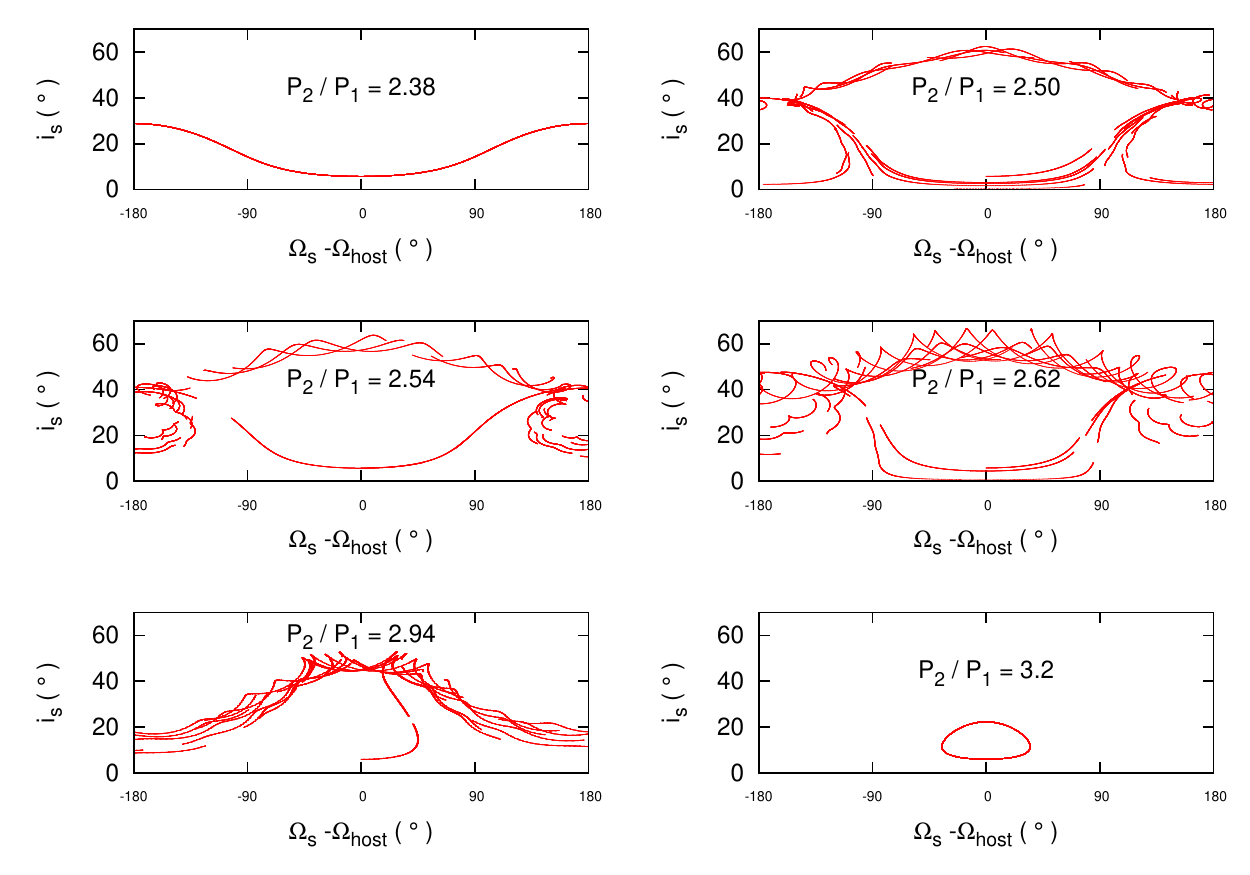}
  \caption{%
Phase space plots of the innermost satellites.  The orbital elements are calculated on the system's invariable plane.  The upper left plot is a case before entering the 3:2 secular resonance showing circulating motion and the lower right a case in the 1:1 resonance showing librating motion. The middle 4 plots are cases in the overlap region of the 3:2 and 1:1 resonance.  This could imply that the transition into the 1:1 resonance is
responsible for the satellite's high inclination.
}\label{fig6}
\end{figure}

In figure \ref{fig6} are the phase space plots of the innermost moons.  The unstable cases show traces of being affected by the 1:1 secular resonance, and they happen in between circulating (upper left) and librating (lower right) motions.  The transition through the separatrices into the 1:1 resonance could likely cause the satellite's inclination to experience large oscillations. 
  And the stability of the innermost satellites in simulation set 2 could be due to the enhanced rate of their nodal precession by the host planet's $J_2$.  In 2 out of 25 simulations where the innermost satellites' precession around the host planet's spin axis occurrs at intervals when it acquires a tiny inclination of 0.02$\degree$, the period is on the order of 10 years.

\subsection{numerical error}

   Simulation set 4 double checks that the behavior of the innermost satellites in set 1 are not attributable to numerical 
error.  The simulation with $\frac{P_2}{P_1}$ = 4.01 in simulation set 1 were rerun with a smaller integration error limit of $10^{-13}$,
 and a shorter integration time step of 0.001 days.  The resulting orbits of the innermost satellites resemble
 those in simulation set 1.  In addition, among all 3 simulation sets, the maximum simulation error in energy dE/E is $2.58 \times 10^{-8}$ and in
 angular momentum dL/L is $7.90 \times 10^{-9}$.  Therefore, we do not attribute the innermost satellites' behavior to numerical error.

\section{Conclusions}

The close-in satellites within the critial distance where planet oblateness usually dominates demonstrate very different dynamics from their Solar System counterparts when the planetary bulge is absent.  The planetary bulge enhances the nodal precession rate of satellites, and without it, the nodal precession rate of a satellite can be slow enough to hit secular resonances with the host planet, which could potentially cause the inclination to increase.  These unusual dynamical effects disappear when the host planet is oblate, as we would expect.  As a side note, the type of instability demonstrated in this work could not occur for perfectly co-planar systems, and is not relevant for satellites beyond $a_{crit}$. 

 Numerical modeling often needs to make assumptions, and it is common to neglect factors not immediately interesting or relevant.  However, when simulating exomoon orbits, we may encounter dynamical systems that have planets on mutually inclined orbits.  Some known exoplanet systems are on more compact and eccentric orbits, 
and they may acquire high mutual inclination as they interact.  They may also encounter each other extremely closely. In those systems, perturbations may often be greater than in this work and without planet oblateness
the system could encounter a similar effect seen in this work.

\section*{Acknowledgments}
We thank Sean Raymond and John Chambers for helpful conversations.  YCH and JIL acknowledge support from the JWST Project.  
MST acknowledges support from NASA Outer Planets Research (NNX10AP94G).

\bibliographystyle{mn2e}
\bibliography{arxiv2}

\label{lastpage}

\end{document}